# Sub-method, partial behavioral reflection with Reflectivity

## Looking back on 10 years of use

Steven Costiou[a], Vincent Aranega[b], and Marcus Denker[a]

a   Inria, Univ. Lille, CNRS, Centrale Lille, UMR 9189 - CRIStAL
b   Univ. Lille, CNRS, Centrale Lille, Inria, UMR 9189 - CRIStAL

**Abstract**   **Context.** Refining or altering existing behavior is the daily work of every developer, but that cannot be always anticipated, and software sometimes cannot be stopped. In such cases, unanticipated adaptation of running systems is of interest for many scenarios, ranging from functional upgrades to *on-the-fly* debugging or monitoring of critical applications.

   **Inquiry.** A way of altering software at run time is using behavioral reflection, which is particularly well-suited for unanticipated adaptation of real-world systems. Partial behavioral reflection is not a new idea, and for years many efforts have been made to propose a practical way of expressing it. Many of these efforts resulted in practical solutions, but which introduced a semantic gap between the code that requires adaptation and the expression of the partial behavior.

   **Approach.** The idea of closing the gap between the code and the expression of the partial behavior led to the implementation of the Reflectivity framework. Using Reflectivity, developers annotate *abstract syntax tree* (AST) nodes with meta-behavior which is taken into account by the compiler to produce behavioral variations. Reflectivity is designed for dynamically typed systems which provide an AST representation of the program that is causally connected to the source code, and which support run-time recompilation. In this paper, we present Reflectivity, its API, its implementation in Pharo, its usage and limitations. We reflect on ten years of use of Reflectivity, and investigate in the literature how it has been used as a basic building block of many innovative ideas.

   **Knowledge.** Reflectivity has been used by 21 projects in the last decade, to implement reflective libraries or language extensions, code instrumentation, dynamic software update, debugging tools and software analysis tools. Our investigation shows that developers needed powerful and customized sets of heterogeneous reflection features for their projects, which Reflectivity provided. Despite its limitations, Reflectivity has proven to be a practical way of working with fine-grained reflective operations (at the AST level), and enabled a powerful way of dynamically add and modify behavior. By instrumenting through AST annotations, Reflectivity provides a flexible means to bridge the gap between the expression of the meta-behavior and the source code.

   **Grounding.** Reflectivity is actively used in research projects. During the past ten years, it served as a support for both implementation and fundamental base, for much research work including PhD theses, papers at conferences, workshops and in journals. Reflectivity is now an important library of the Pharo language, and is integrated in the standard distribution of the platform.

   **Importance.** Reflectivity exposes powerful abstractions to deal with behavioral adaptation, while providing a mature framework for unanticipated, non-intrusive, sub-method and partial behavioral reflection based on AST annotation. Finally, the AST annotation feature of Reflectivity opens new experimentation opportunities about the control that developers could gain on the behavior of their own software.

# The Art, Science, and Engineering of Programming

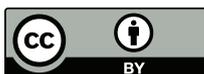

**Sub-method, partial behavioral reflection with Reflectivity**

## 1 Introduction

A system is called reflective if it is able to reason about itself and if it embeds a model of itself in a way that it is causally connected [31, 53]. Changing the model changes the system itself, and any change to the system is reflected in the model [22]. Reflection[1] [6, 15, 29, 30, 37, 46, 52, 53, 55, 61] provides the basis for a class of features for various language capabilities and tool implementation. As examples we can name run-time adaptation [44], debugging complex embedded systems in deployment [11] or tools that allow developers to explore and understand a system at run time [64].

For thinking about reflection, it makes often sense to look at two distinct (but connected) points of view: structure and behavior [29]. *Structural reflection* is concerned with modeling the structure of a system [6]. Concepts of the language, in the case of an object-oriented system, for example classes, methods, packages are represented as objects inside the language at run time. It can be limited to observation (i.e., read-only) or subject to change. Changing this representation changes the system itself. We can add, remove or change classes or methods and the change is active in the currently running system. *Behavioral reflection*, also named *intercession* [4], is concerned with run-time capabilities, for example intercepting message sends and receives, instance variable reads and writes or reasoning about control flow [34].

In general purpose languages, reflection is limited as it is challenging to create a system that is fully reflective and practical in terms of execution speed and memory consumption. Even languages like Smalltalk that have extensive reflective features present many possibilities for improvement [45].

In the case of Smalltalk, structural reflection stops at the method level: methods are objects, but the structure inside of methods is not modeled. From a behavioral point of view, neither message sending nor access to variables are reflectively accessible. The Reflectivity subsystem of Pharo [3] is a practical approach to improve the reflective capabilities of Pharo from both a structural and behavioral point of view.

Behavioral reflection using Reflectivity, paired with the live programming paradigm provided by Smalltalk, provides a stable and robust way of dealing with unanticipated behavioral adaptation. Reflectivity has been put to the use in many contexts, either in Pharo itself, where it provides key mechanisms for the system, or in connected works.

In this paper, we look back on 10 years of Reflectivity in Pharo and show how sub-method, unanticipated behavioral reflection has been used. We first present works related to behavioral reflection for object-oriented languages in section 2. We then present the fundamental properties of Reflectivity in section 3, whose use is illustrated by examples in section 4. In section 5, we discuss through a critical analysis the reasons for which Reflectivity was used in different projects over the years, how its features were used and the limitations of the system. Finally in section 6 we present future directions for Reflectivity. In addition, we provide in appendix B an overview of projects which used Reflectivity. In appendix C we provide a report of a performance analysis performed on the latest implementation of Reflectivity.

---

[1] A deeper study of reflection and its implementation has been done by Tanter [59].

## 2 Related Work

In this section, we study similar systems providing partial behavioral reflection that is unanticipated and applicable to the sub-method level. We describe work which inspired Reflectivity (section 2.1) and closely related work (section 2.2 and 2.3). We then describe recent work on efficient and dynamic AST instrumentation (section 2.4) as well as work on sub-method reflection (section 2.5).

### 2.1 Where Reflectivity comes from

Reflectivity originally comes from the combination of *Unanticipated Partial Behavioral Reflection* [48] and *Sub-Method Reflection* [17]. These concepts which led to the current version of Reflectivity are described through the following history. We note that the following prior work discusses itself related work in detail, we therefore do not repeat such discussion in this paper.

This paper thus is not only a report of the usage of Reflectivity over the last 10 years, but in addition it presents the full model for the first time. We end this section with a short discussion detailing the delta to prior publications.

**Partial Behavioral Reflection.** *Reflex* [61] is a Java implementation of *PBR*, a fine-grained Meta-Object Protocol (or MOP, an interface to control meta-objects [29]). *PBR* provides fine-grained spatial and temporal selection over controlled behavior and reified run-time entities. Reflex pioneers the concept of metalinks. But here, the link is installed on a hook set. A hook set defines a set of operations at the Java bytecode level. Hooksets can only be defined at load time and thus have to be anticipated by the programmer, while the (un)installation of links happens dynamically at run time.

As Reflex was working with bytecode level abstractions, it was realized using bytecode transformation. We discuss related work in section 2.2.

Partial Behavioral Reflection can serve as an implementation substrate for Aspect Oriented Programming [27]. A discussion of the relationship between AOP and Partial Behavioral Reflection is out of the scope of this paper and has been discussed by Tanter [62].

**Unanticipated Partial Behavioral Reflection.** *UPBR*, implemented by *Geppetto* [48] allows for the definition and retraction of reflective features at run time. *UPBR* reimplements *PBR* in Squeak [26] using bytecode manipulation and rewriting, following closely the implementation of *Reflex*.

The capability to change bytecode at run time leads to a much more flexible model: Partial Behavioral Reflection can be done without anticipating at load time where it might be used later at run time.

**Sub-Method Reflection.** *SMR*, implemented as *Persephone* [17], refines the granularity of the structural model to sub-elements of methods in a way that it is causally connected: changing the model changes the system. To that end, *SMR* introduced *Dual Methods*, in which compiled methods have a twin method named *reflective method*.

**Sub-method, partial behavioral reflection with Reflectivity**

This *reflective method* is the meta-object of the compiled method it is associated with, and provides access to the AST of this compiled method. Every compiled method of the program is transformed into a dual method at compile time. *Persephone* supports annotations (textual and non-textual). Compiler plugins can be defined to give meaning to annotations with regard to execution. It is basically a *sub-method* MOP, based on AST annotations. *Persephone* was originally implemented in Squeak [26].

**Contributions compared to prior publications.** Reflectivity is a implementation of *Unanticipated Partial Behavioral Reflection* [48] and of *Sub-Method Reflection* [17], both combined in a single model with the AST replacing the bytecode level hookset. This idea was first formulated in the later chapters of Denker's thesis [16]. This paper presents the full model for the first time.

From an implementation standpoint, the current version is based on a completely redesigned *Persephone* [17], to make creation of *dual methods* available on demand instead of systematically creating a *reflective method* for every compiled method of the program. *Reflective methods* are only created when meta-behavior is installed on an AST node of a method, and automatically removed when the last annotated AST is removed from this method. Reflectivity implements meta levels (section 2.3) as described in our prior publication [20].

Reflectivity has been extended to support object-centric reflection. Metalinks can now be scoped to specific objects, we discuss this contribution in section 3.4.

**2.2 Bytecode Transformation**

Both Reflex and our implementation of Reflex in Smalltalk are realized using bytecode transformation. Reflex uses Javasist [6, 8], the Geppetto implementation of Unanticipated Partial behavioral reflection that was realized with Bytesurgeon [18]. We refer to the publication about Bytesurgeon for a discussion of related work.

Making different independent bytecode instrumentation coexist over a same original base code requires significant engineering effort. Moret *et al.,* implemented polymorphic bytecode intrumentation as the process of composing multiple independent instrumentation of a class and selecting at run time which version to execute [32, 36]. The different bytecode instrumentations are applied to an original class producing different instrumented versions of the class. Those versions are merged together in a final class along with the dynamic dispatched logic defined by the user. Instrumentation of the original class is performed at load time, imposes some restriction on the instrumentation and the granularity of the instrumentation is at the method level.

In contrast, Reflectivity does not produce a new instrumented class. Instead, it creates a reflective method keeping a relationship towards the original method. This makes possible the (un)installation of the instrumentation at run time. Polymorphic bytecode instrumentation is a bytecode level framework while Reflectivity focuses on AST annotation and reflective methods, regardless of how it is implemented.

Tools that rely on bytecode instrumentation are tedious to develop either because of the use of low-level bytecode manipulation libraries or because of the AOP join point model. That model offers a higher level of abstraction than bytecode, but is not suited for many tasks that require to deal with different fine-grained join points (e.g., the execution of bytecode or basic blocks). To overcome these limitations, Marek *et al.,* proposed DiSL [32], a language designed to *shadow* any region of the bytecode (i.e., to define any region of the bytecode as join point [23]) and to access static and dynamic context information as reifications [32]. However, it does not provide any capability for scoping instrumented code to specific objects. The language uses Java annotations and a set of *Markers* to express which part of the bytecode should be defined as a shadow. *Guards* and *Scopes* provide two complementary mechanisms for restricting the application of snippets (i.e., a static method in an instrumentation class) at weave time.

Unlike Reflectivity where the AST is annotated by metalinks at run time, the Java annotations to define new joint points are set in the code over methods definitions at weave time, and cannot be applied in an unanticipated manner.

### 2.3 Taming endless meta recursion

When performing reflective calls to meta objects from system level code, it can happen that the meta-object calls the same method again, leading to endless recursion.

The original model of reflection as defined by Smith [53] is based on meta-level interpretation. The program is interpreted by an interpreter, this interpreter is interpreted by a metainterpreter leading to a tower of interpreters each defining the semantics of the program or interpreter it interprets. There is no problem of meta recursion, because each interpreter forms a context that defines if we are executing at the base level or at the meta-level. Calling reflective functionality is always specific to one interpreter. The meta-object approach in contrast is lacking any mechanism to specify this contextual information.

In the context of Reflectivity, the meta-object based reflective system is extended with a first class notion of meta-level execution, and the possibility to query at any point in the execution if we are executing at the meta-level [20]. To model meta-level execution we propose the notion of a first class context and context-aware reifications. Reflectivity allows the programmer to define a *level* on which a link is active. The system itself executes on level 0, every link activation increments this level, and returning from the meta-level code decreases it. For solving the recursion problem, two levels would be enough, as this would allow to distinguish base from meta execution. A level modeled as an integer is more general. We can, for example, define links to be only active on the meta level to analyse meta-level execution itself.

Chiba discusses this problem in the context of reflectively defining new kinds of slots in Common Lisp [7]. The focus is the problem that fields added by reflection to implement changed behavior show through to any user of introspection. The solution (the *MetaHelix*) is both more general and restrictive than context based solutions. It is more general, as it tries to solve the problem of the visibility of structural change. And it is more restrictive, as it does not model meta-level execution and instead relies on the programmer to structure the meta level code to avoid recursion. A detailed discussion of the *MetaHelix* and other related work can be found in our prior publication [20].

**Sub-method, partial behavioral reflection with Reflectivity**

Tanter discussed execution levels in the context of AOP [60]. Polymorphic bytecode instrumentation [36] has been used for implementing execution levels for AspectJ. To count infinite recursion in general, Polymorphic bytecode instrumentation implements a simpler control flow check as a boolean value per thread of execution. This can be seen as a simplified level model (with two levels: base and meta) targeted to specifically solve the recursion problem.

## 2.4 Dynamic and optimized AST instrumentation

The Truffle Instrumentation Framework [66] is an extension of the GraalVM[2] virtual machine, for dynamic, efficient and non-intrusive program instrumentation at the level of the AST. The GraalVM is a virtual machine that runs programs written in Truffle-based languages [69, 70]. Those programs are AST based, and are executed by an interpreter which dynamically specializes and optimizes nodes of programs' ASTs. The GraalVM then dynamically optimizes the Truffle interpreter for that AST into efficient compiled code. The Truffle Instrumentation Framework [66] inserts instrumentation nodes into the AST of a *Truffle-language* program, producing high-performance instrumented runtimes.

Such instrumentation of the program's AST is a means of achieving sub-method reflection, and to a given extent partial behavioral-reflection. However, it seems not possible yet to fully replace an original node by another node (the instead of *Reflex* and Reflectivity), nor to query specific part of the execution context (the reifications). Object-centric instrumentation, although it seems possible through the conditioning of the instrumentation, is not covered by the Truffle Instrumentation Framework. However, the possibilities of efficient and flexible instrumentation offered by the framework are appealing for experimenting new and high-performance implementations of Reflectivity in Truffle based languages.

## 2.5 Sub-method reflection

In the following, we present works related to reflection at the sub-method level.

Annotations are a simple mechanism that allows one to introduce meta-information directly in the code. They are used by Java since Java 1.5 and some Smalltalk dialects like Squeak, VisualWork and Pharo. Spoon is an open compiler for Java that provides compile-time reflection [40]. It provides support for the Java annotations (i.e., the transformation processor can read annotations) and gives access to the Java AST that can be transformed before it is compiled to bytecode. Spoon works at compile time, not run time. Moreover, the AST is not available at run time, and the compile-time framework is limited to the annotation model provided by Java.

We can find other mechanisms to give access to method structural reflection as in LISP [54]. KSL represents all languages constructs as objects [25]. Its way of working is the closer to our approach as it provides access to sub-method elements as objects at run time. But KSL is purely interpreted, the represention is not compiled, not extensible and not used to provide behavioral reflection.

---

[2] http://www.graalvm.org/.

## 3 Reflectivity: Overview

Reflectivity improves the reflective abilities concerning both structure and behavior [16]. The Reflectivity model therefore consists of two related parts: *Sub-method reflection* [17] and *Partial Behavioral Reflection* [61].

**Sub-method Reflection.** It provides a fine-grained model of method structure down to the AST level, by making the AST available inside the language [17].

**Partial Behavioral Reflection.** On top of the structure provided by sub-method reflection, Reflectivity implements behavioral reflection by annotating the structural model with *metalinks*. To that end, a metalink annotates an AST node, that is expanded (with instrumentation), compiled and executed *on-the-fly*. A metalink can be seen as being installed on the base-level of a program, leading to execution of code on a meta-level. An overview of AST annotation with metalinks can be seen in figure 1.

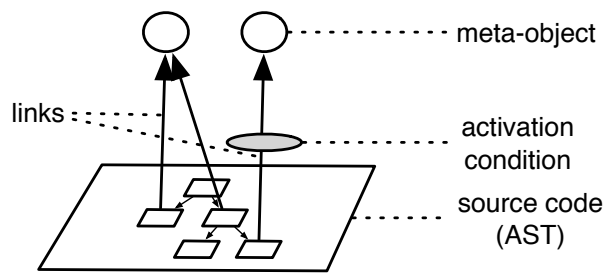

**Figure 1** Metalinks installed on the AST

### 3.1 The Metalink

Metalinks are the core entities which provide control over reflective behavior, exposed through an API. A metalink annotates one or many nodes of the AST of a method. Such annotation defines a reflective action to perform, a meta-object on which performing this action, and a condition. When the program execution reaches the annotated AST in the control flow, the metalink is *executed*. A metalink is then composed of the following elements:

**The meta-object** is an object implementing the meta-behavior that is executed when the link is triggered.

**The selector** is the message selector that is sent to the meta-object when the link is triggered, resulting in the execution of the corresponding method in the receiver (i.e., the meta-object).

**The arguments** specify what information is passed to the meta level.

**The control** specifies when the metalink is triggered, i.e., the moment when the message corresponding to the given selector will be sent to the *meta-object*. It can be before, instead or after the execution of the operation defined by the AST node on which the metalink is installed.

**The condition** is either a boolean or a block closure (first class anonymous function) evaluating to a boolean. The condition is checked at each activation of a link and control is only passed to the meta-object if the condition evaluates to true.

A Metalink is a first-class object, that is instantiated, configured (by setting the aforementioned elements) and installed by its clients (developers or tools). Metalinks are mutable, when installed metalinks are modified (for exampe a condition is added), the changed definition is installed without the need to explicitly reinstall the link.

### 3.2 Reifications

The metalink defines a message send from base level code to the meta-object. For example, a link on a message send might want to pass the receiver or arguments to the meta object. Reflectivity provides list of entities that can be reified from the execution context. Which entities should be reified is specified by configuring the metalink. Reified execution entities can be used by the metalink as arguments passed either to the meta-object or to the conditional block that guards the meta-behavior execution. Reifications themselves can even be used as meta-objects by metalinks. The full list of reifications is available in table 3 from appendix A.

### 3.3 The use of the AST

The metalink model originally comes from Reflex where it was realized on a bytecode-level representation: operations on the bytecode level form a hookset. The metalink is then installed on a hookset. With Reflectivity we took the metalink but we applied it to an AST based reflective model of code. In our prior work about *Sub-Method Reflection* [17], we analyse the challenges for sub-method reflection using text, bytecode and the AST, and show how the AST can be used as a solution without the problems that the other representations present.

The most important is what one can call the semantic gap: bytecode is optimized for execution, not reflection. In the case of Pharo, this goes so far that some information cannot be recovered, as some code is compiled to optimized bytecode, for example.

Deeper discussion on why the AST is an interesting representation for sub-method structure can be found in our prior work [17].

### 3.4 Object-centric reflection

A metalink can be scoped to a single, specific object. In that case, the meta-behavior specified by the metalink is only executed for that particular object, and all other instances of the program are unaffected. Class-wide metalinks (i.e., metalinks affecting all instances of a class) and object-centric metalinks can be composed. The same AST node can be annotated with class-wide metalinks and object-centric metalinks only affecting specific instances. An object-centric metalink can be installed on many different objects, each of these objects becoming individually affected by the metalink. The metalink can then be either selectively removed from one or more of these objects, or entirely uninstalled from all objects at once.

### 3.5 Meta-Level Recursion

Endless meta recursion can happen when installing a metalink in system level code. This is especially a problem when dealing with intensively used parts of the system, for example Numbers, Strings, Collections or the language kernel itself. To solve this problem, Reflectivity implements meta levels [20]. Metalinks are configured with an explicit level (an integer) to control its (de)activation and prevent meta recursion.

### 3.6 Properties

In the following, we describe the different properties of Reflectivity.

**Partiality.** The metalink model leads to partial behavioral reflection such as defined in section 2. Reflection can be selected both spatially and temporally:

**Temporal selection** The metalink can be installed and retracted at run time. In addition, the condition can be used to control *metalinks* taking additional run-time information into account.

**Spatial selection** We can select precisely, down to a single operation, where we want to install a metalink.

**Selective reification** A metalink does not pass all information to the meta-level, but defines exactly which information (if any) to pass to the meta-object.

**Cross-cutting.** Cross-cutting is a term from *Aspect Oriented Programming* [27]. The standard example is logging: the code to log program activity (e.g., for debugging) is not modular in one place, but spread over many classes, methods and packages. Aspects allow the programmer to implement cross-cutting concerns in one place.

With Reflectivity we can do the same, but from the reflective point of view: a single metalink can be installed for one operation cross-cutting many class hierarchies. It can even cross-cut from an operational point of view. We can define a metalink and install it on many AST nodes, even if these nodes are representing different operations. We can, for example, install the same metalink on both instance variable access and message sends. This allows for one meta-object to cross-cut over different operations spanning multiple class hierarchies.

**Compilable.** The Reflectivity model with the AST and *metalinks* as annotations can be implemented efficiently by transforming and compiling the annotated AST before execution. This implies that the running program includes a compiler, which is the case of programs deployed on the standard Pharo distribution. Systems where there is no compiler at run time will only support instrumentation before run time, or rely

**Sub-method, partial behavioral reflection with Reflectivity**

on a hot code update mechanism. Hot update techniques are nowadays available for many languages such as, for example, C [38], Java [41] or Pharo [63], as well as for a large variety of systems [2, 5, 24, 57, 67, 68].

### 3.7 Implementation needs

The Reflectivity API is built on top of three main concepts: meta-objects, *metalinks* and AST nodes. One important design decision guiding the implementation was not to be based on a modified virtual machine. This makes the system available to all users of Pharo without the need to maintain a modified virtual machine. This in addition means that Reflectivity can be implemented in any language implementation that is able to:

1. expose the program's AST,
2. dynamically compile a new method from an instrumented AST,
3. replace a method by another one at run time.

In this paper, we present a Pharo implementation, while a Python implementation is available online.[3]

### 3.8 Pharo Implementation Aspects

Reflectivity achieves sub-method and partial behavioral reflection by the instantiation, the configuration and the installation of *metalinks* on AST nodes of a program. The current Pharo implementation of Reflectivity takes advantage of the Pharo live programming environment, and makes possible to define and modify reflective operations in an unanticipated manner while the program is running.

**Sub-method reflection.** Reflectivity reimplements sub-method reflection by providing on-demand surrogates of instrumented methods named *reflective-methods* as sketched in figure 2.

Each time a metalink is installed on a method, the AST of this original method is annotated. A *reflective-method* is generated by recompiling the code from the original method with instrumentation defined by *metalinks* installed on its AST. This reflective method is then associated with this original method, and replaces it at execution time. When debugging, e.g., when a breakpoint halts the program, it is the source code of the original method that is displayed and navigated through debugging operations.

Even though the original source is shown, when executing code (including stepping in the debugger), the code generated for the metalink is executed. The IDE can help with visualizing metalinks as well as providing advanced debug support. The current version of Reflectivity does not provide IDE support for this yet.

---

[3] https://github.com/StevenCostiou/reflectivipy.

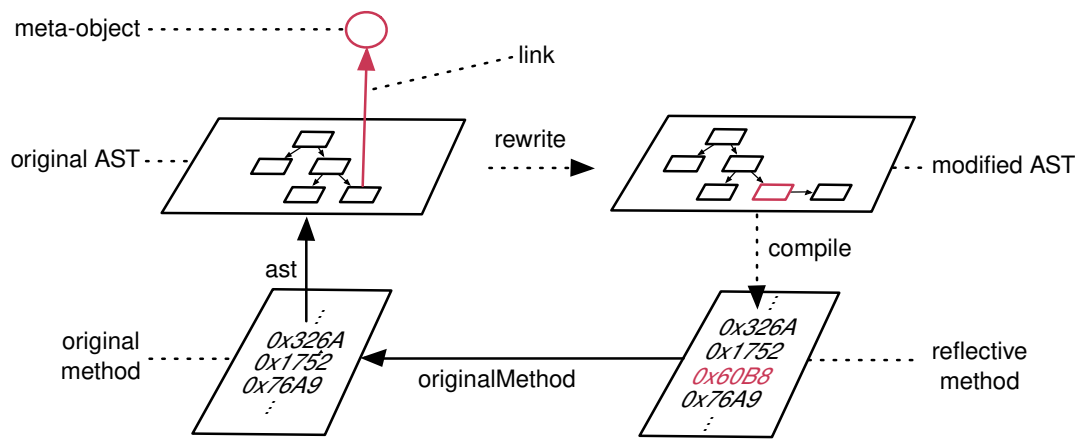

**Figure 2** Sub-method Reflection, *Original* and *Reflective* methods

## 4 Some Reflectivity Examples

To install a metalink on an AST node inside a method, the developer has to navigate the method's AST. This can be done either programmatically, or by using the integrated tools of the environment. This can be done by obtaining the method AST object (like in listing 2, line 1) and programmatically navigate into this AST to find the node we are interested in. The API can then be used to install the metalink on that particular node. It can also be done through the contextual menu of the Pharo class browser. This is illustrated in figure 3, which shows the code of the `logCr` method in the class browser. We selected one of the message sends in the method, and used the contextual menu to select the metalink (instantiated in listing 1) and to install it on this specific node. The execution will only execute the metalink when the selected code (highlighted in blue) will be reached in the control flow.

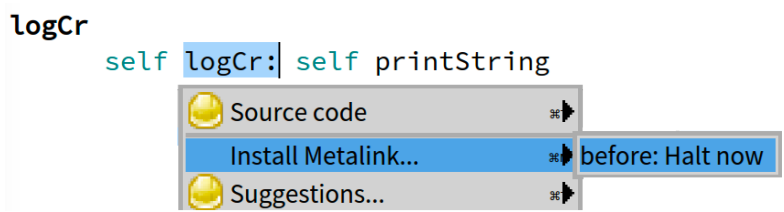

**Figure 3** Integrated AST node selection for MetaLink installation in the Pharo class browser

### 4.1 Example: breakpoints with metalinks.

In listing 1, we illustrate a basic example of Reflectivity with a breakpoint implementation. A `MetaLink` is first instantiated (line 1). The metalink is configured with a meta-object (line 2), a selector (line 3) and a control (line 4).

**Listing 1** MetaLink breakpoint example

```
1  metalink := MetaLink new.
2  metalink metaObject: Halt.
```

**Sub-method, partial behavioral reflection with Reflectivity**

```
3  metalink selector: #now.
4  metalink control: #before.
```

Once a metalink has been configured, it can be installed on any AST node. This is illustrated in listing 2. To install a metalink, one must first access an AST node. In Pharo, it can be done sending the #lookupSelector: message to a class, with a given selector, which will return the corresponding compiled method. The AST can then be asked to the method object. In listing 2, we recover the method node from the logCr method of the Object class (line 1). The metalink can be installed on this node by sending the #link: message to the node and passing the metalink as a parameter (line 2). After that, each time the #logCr message will be received by an instance of Object, as at line 3, the metalink will be triggered before the execution of the logCr method and send the #now message to the Halt class (i.e., breaking the execution). A metalink can be installed on many different nodes at the same time, without any limit.

■ **Listing 2**  MetaLink installation example

```
1  node := (Object lookupSelector: #logCr) ast. "Getting the root AST node of the logCr method"
2  node link: metalink.  "Annotating the node with the metalink"
3  Object new logCr.
```

Finally, listing 3 shows how metalinks can be uninstalled. A metalink can be either removed from a specific node (line 1) or completely uninstalled (line 2). In that latter case, the metalink is uninstalled from every node it was installed on.

■ **Listing 3**  MetaLink uninstallation example

```
1  node removeLink: metalink.
2  metalink uninstall
```

### 4.2  Example: Partial behavioral reflection.

This example shows all three aspects of partiality. We install a metalink on just one operation. Then we pass selected reifications as an array of arguments. This array contains symbols specifying which aspect of the execution should be reified. Finally, we show how to use conditions to control link activation.

Listing 4 shows the configuration of a metalink with reifications from the execution context. After its instantiation (line 1), we request two reifications from the metalink: receiver and arguments (line 4). These reifications are intended to be respectively the receiver of a message and the arguments passed to this message in the control flow (see table 3 from appendix A for more details). At this point, it is implicit that the node on which the metalink will be installed has to be an AST message-send node, because these reifications are only available for this node.

In this example, a block is passed as the meta-object (line 7). This block takes two parameters, receiver and arguments, which are the reifications previously specified. In the block, we define meta-behavior using the arguments of this block. Here, a class browser will be opened on the class of the receiver (line 8), and an inspector will be opened on the receiver and on the arguments (lines 9-10).

A selector is configured as a message selector taking two arguments (#value:value:) that will be sent to the meta-object when the metalink executes. At run time, the two reifications requested at line 4 will be passed to this selector as parameters and the corresponding message will be sent to the block (the meta-object), resulting to the execution of the meta-behavior defined by the block.

The metalink is installed on the #logCr: method of the Object class (lines 17-18). We instantiate an OrderedCollection, and add two random numbers in it (lines 20-22). Then we send the #logCr: message to the collection to print its size (line 26), and the meta-behavior is triggered.

The two arguments passed to the meta-object are the receiver of the #logCr: message (i.e., the ordered collection object) and its arguments (an array). The meta-behavior opens a class browser on the OrderedCollection class, and inspectors on the collection object and on the arguments of the #logCr: message.

■ **Listing 4** Reifications of contextual entities with MetaLink

```
1  metalink := MetaLink new.
2
3  "Reification requests: an array of symbols referring to run-time entities"
4  metalink arguments: #(receiver arguments).
5
6  "The meta-object is a block taking the reifications configured above as arguments"
7  metalink metaObject: [:receiver :arguments|
8                              receiver class browse.
9                              receiver inspect.
10                             arguments inspect.].
11
12 "Selector that will be sent to the meta-object"
13 metalink selector: #value:value:.
14 metalink control: #before.
15
16 "Getting the root AST node of the logCr: method, on which we install the metalink"
17 node := (Object lookupSelector: #logCr:) ast.
18 node link: metalink.
19
20 collection := OrderedCollection new.
21 collection add: Random new next.
22 collection add: Random new next.
23
24 "The metalink triggers before the logCr: method, and executes the meta-object
25 with as parameters the receiver (collection) and the arguments of the message send"
26 collection logCr: 'Size: ', collection size printString.
```

To illustrate conditional metalinks, we reuse the class and the collection instantiated in listing 4. Metalinks are made conditional through the #condition:arguments: message. The parameters to this message are a block defining the condition and an array of arguments that will be passed to the block at execution time. For instance, we define in listing 5 a conditional block with one parameter and we specify that this parameter is the receiver of the message being executed (line 2). The condition tests if the receiver (i.e., the collection object from listing 4) contains more than two elements. Continuing in listing 5, the invalidation of the metalink (line 2) updates the instrumented code.

**Sub-method, partial behavioral reflection with Reflectivity**

When the metalink is triggered, i.e., the node on which it is installed is executed (the #logCr: message line 3), the condition is first evaluated and the meta-behavior specified by the metalink is executed only if the condition is evaluated to true.

■ **Listing 5**   MetaLink condition using reifications

```
1  "Conditions take a block with parameters, which are reifications from the execution context"
2  metalink condition: [:receiver| receiver size > 2] arguments: #(receiver).
3
4  metalink invalidate.
5  collection logCr: 'Size: ', collection size printString.
```

### 4.3  Example: object-centric breakpoints with MetaLink.

Object-centric breakpoints can immediately be implemented using object-centric metalinks. The metalink from listing 1 can be installed on a single object to provide an object-centric breakpoint, which is illustrated in listing 6. First, the breakpoint metalink is instantiated like in listing 1 (line 1), then two instances of OrderedCollection are created (line 2-3). The AST node of the logCr: method is obtained (line 6) and the metalink is installed on this node specifically for the collection_1 object (line 9). The target object on which installing the metalink is specified through the #link:forObject: interface. The execution will break each time the collection_1 object will receive the #logCr: (line 11), while other instances of OrderedCollection will not (like collection_2, line 12).

■ **Listing 6**   MetaLink object-centric breakpoint example

```
1  metalink := "...Breakpoint metalink..."
2  collection_1 := OrderedCollection new.
3  collection_2 := OrderedCollection new.
4
5  "Obtaining the AST node of the logCr method"
6  node := (Object lookupSelector: #logCr:) ast.
7
8  "Installation of the metalink on the collection_1 object"
9  node link: metalink forObject: collection_1.
10
11 collection_1 logCr: 'Break'
12 collection_2 logCr: 'No break'
```

### 4.4  Example: Meta-Level Recursion

A metalink can be configured to be active only at a certain level by setting the level when configuring the metalink:

■ **Listing 7**   MetaLink level example

```
1  metalink := MetaLink new.
2  metalink metaObject: Halt.
3  metalink selector: #now.
4  metalink control: #before.
```

```
5
6  "The meta-level is configured to 0: the breakpoint will trigger at the base level
7  and will stop triggering when executing code from the meta-level"
8  metalink level: 0.
```

This metalink will only be active when called on the base level of the program. When the control flow reaches the metalink from the meta-object code itself (in this case the execution of now in class Halt), the metalink will not be active.

## 5 Why and how Reflectivity is used

In this section, we develop a critical analysis of Reflectivity's usage in 21 different works. We discuss what researchers and developers were looking for when they chose to use Reflectivity for their project. Then, we discuss the use-cases for Reflectivity's features and how they are used in real projects. Finally, we discuss limitations of Reflectivity in the light of the usage that has been made in the last ten years.

### 5.1 Why Reflectivity

In most of the work we studied, Reflectivity is chosen when developers look for reflection capabilities providing means to dynamically instrument programs, with the finest possible granularity, and with minimal impact on programs' source code. Reflectivity fulfills these three requirements, through unanticipated reflection, sub-method, partial behavioral reflection, and non-intrusive instrumentation. Through the analysis of the aforementioned works, we report in the following why these three features and their combination were required.

**Sub-method and partial behavioral reflection.** Reflectivity combines sub-method [17] and partial behavioral reflection [48] . First, sub-method reflection enables fine grained reflection, at the AST level. This granularity is required by solutions and tools providing behavioral instrumentation within the body of methods, *i.e.*, which only apply reflection to restricted parts of a method's body [1, 9, 10, 21, 56, 64, 65]. Second, partial behavioral reflection provides means to specify when reflection will be triggered, and which entities of the execution context will be reified. This enables fine and customized scoping of the reified context [19, 44, 47, 49, 50, 51, 64, 65]. Finally, the combination of sub-method and partial behavioral reflection is needed to define and apply controlled and fine grained behavioral variations. For instance, object-centric debuggers [11, 12, 13, 14] need to capture very specific objects from the execution context, and to insert instrumentation at strategic places in their methods for debugging.

**Unanticipated reflection.** We define *unanticipated reflection* as reflective operations that are unknown and unforeseen until they are needed at run time. Therefore, these operations must be specified and applied on-demand to the running program. Reflectivity's features can be applied in such an unanticipated way, which is a reason for one to choose Reflectivity as an implementation layer [19, 44]. That capability is

**Sub-method, partial behavioral reflection with Reflectivity**

fundamental to projects aiming to dynamically change and/or to control behavior of a running program [11, 12, 13, 14, 19, 43, 44], to enable/disable reflective features *on-the-fly* [56], or to provide on-demand, instant access to dynamic data [47, 49, 50, 51].

**Non-intrusive instrumentation.**   Reflectivity instrumentations are not inserted in the base code, but work from the meta-level. This maintains a strong and clear separation between instrumentation and program behavior. Instrumentation can be added or removed without modifying the code of the program or system. The root reason for which developers look for this non-intrusiveness property is that no additional (non-functional) statement has to be inserted in their program. This is exploited by developers, e.g., to avoid code versioning noise [1] (because code has been modified with instrumentation), to separate debugging behavior and debugged programs' behavior [11, 13], to intercept and profile meta-events from the Pharo IDE without modifying its source code [35], or to cope with constraints on code modifications [65]. To visualize and control instrumentation, which is invisible to users and to system tools, developers often built and used dedicated tools suiting their own usecase [1, 11, 12, 14, 64, 65].

## 5.2 Use-cases for Reflectivity's feature usage

Most of the studied projects use an heterogeneous and customized combination of Reflectivity's features as an implementation layer. Some projects only exploit Reflectivity for a single particular feature. While the rationale for choosing Reflectivity is described by the analysis in section 5.1, in this section we try to characterize the precise feature set used by each project. In addition, we illustrate concrete usage of those features, by describing how they are used in a selection of four projects.

Table 1 summarizes Reflectivity's features, as they are described in section 3. Each feature is tagged with a letter. Table 2 summarizes the studied projects regrouped by topic. For each project, we summarized the general usage that is made of Reflectivity's features, along with the features used by those projects (tags presented in table 1).

**Reflectivity as a reflection layer.**   It is difficult to extract a rationale from the different and heterogeneous combinations of Reflectivity's features. Besides inserting instrumentation (i) which is used by every project, the most common combination is the selective reifications (r) and the spatial selection (s). This combination is used across all kind of projects. We can see that projects building language libraries or tools (e.g., debugging) exploit a large set of the features, if not all. Projects with specific reflection needs, like code instrumentation, visualizations or very specific debugging tools, use a customized set of the features. Except the two projects using all or almost all of the features, every project use its own unique feature combination.

We believe this shows the feature set is rich enough for Reflectivity to be used as a general reflection layer, from which users can choose and compose the reflection capabilities they need for their project. Of course, we can only report the feature

■ **Table 1** Reflectivity features overview.

| Tag | Feature | | |
|---|---|---|---|
| i | simple instrumentation (metalink) | o | object-centric metalinks |
| cont | metalink control | du | dynamic (un)instrumentation |
| cond | metalink condition | s | spatial selection |
| el | metalink execution level control | ni | non-intrusive instrumentation |
| r | selective reifications | cc | cross-cutting instrumentation |

usage from what was reported in the work we studied. It is possible some feature usage was not reported because of an impact of limited importance.

**How are the features used? An illustration.** We selected four projects from table 2 which cover all the features from table 1. For each project, we selected the subset of features that is the most relevant in their Reflectivity usage. While describing each feature usage, we added its corresponding feature tag from table 1.

- **Dynamic Aspects [56].** In AOP [28], developers inject cross-cutting behavior (*advices*) at key points in the base code (*pointcuts*). *Aspects* are modeled by metalink instances, and their meta-object is used to model advices (i). Pointcuts are expressed through a high level API, which translates pointcuts into a set of AST nodes (s) that are annotated by the same aspect metalink (cc). The API uses reifications (r) as run-time arguments in conditions to control aspect execution. The dynamic property of aspects is immediate as metalinks can be (un)installed at run time (du).
- **Breakpoint support for Pharo [1].** Traditional Smalltalk breakpoints are self halt statements manually inserted in the program. This pollutes source code, interferes with code versioning, and triggers infinite recursions when the instrumented code calls itself. In current Pharo implementation, a breakpoint configures a metalink and a meta-object with a self halt instruction (i). At run time, the meta-object executes this instruction, which breaks the execution. Through a menu in the class browser, users select an expression in the source code, to which is mapped a single AST node (s). The breakpoint metalink is then installed to be triggered before the execution of this node (cont). The execution breaking instruction (the self halt) is not inserted in the source code (ni) as it is called from the metalink. Infinite recursions are solved by configuring the metalink not to trigger itself recursively (el).
- **The Moldable debugger [9, 10].** The Moldable Debugger is a framework to build domain-specific debuggers. A debugger has been extended with control over breakpoint activation for method calls and message sends. One single custom metalink (cc) is installed on a selection of AST nodes (s), corresponding to method calls (the root AST node of a method) or to message sends (all nodes within a method that represent a call to a target method). Metalinks are triggered before (cont) the execution of the node on which they are installed, and the meta-object checks

**Table 2** Reflectivity usage in research projects.

| Research project | Description | i | cont | cond | el | r | o | du | s | ni | cc |
|---|---|---|---|---|---|---|---|---|---|---|---|
| Run-time evolution[44] | Selective reflection triggering and metadata reification to reify and control method lookup in a MOP implementation. | ✓ | ✗ | ✗ | ✗ | ✓ | ✗ | ✓ | ✓ | ✗ | ✗ |
| Profiling IDE interactions[35] | Non-intrusive profiling of meta messages (such as #inspect) and their meta data (e.g., receivers, arguments...) before reception by an object in the IDE. Profiled data are used to analyze how developers interact with their IDE. | ✓ | ✓ | ✗ | ✗ | ✓ | ✗ | ✓ | ✓ | ✓ | ✓ |
| Dynamic synchronization[43] | Dynamic & non-intrusive code adaptation of methods with synchronization behavior (mutexes). Separates synchronization code from functional code. | ✓ | ✗ | ✗ | ✗ | ✗ | ✗ | ✗ | ✗ | ✓ | ✗ |
| IDE enhancement[47, 49, 50, 51] | Metadata capture from executions of specified methods (e.g., receiver & sender of a message, var. name & value of an assignment). These data are integrated to the IDE to provide run-time information to developers. | ✓ | ✗ | ✗ | ✗ | ✓ | ✗ | ✓ | ✓ | ✓ | ✗ |
| Breakpoint support[1] | Non-intrusive breakpoints at the level of sub-expressions. These breakpoints control reflective calls to avoid infinite recursions. | ✓ | ✓ | ✓ | ✓ | ✗ | ✗ | ✗ | ✓ | ✓ | ✗ |
| Remote debugging[33] | Interception of messages sent to (remote) file objects, and dynamically replacing them by calls to proxies performing remote access to the real files. | ✓ | ✗ | ✗ | ✗ | ✗ | ✗ | ✓ | ✓ | ✗ | ✗ |
| On-the-fly* debugging [11, 12, 13, 14] | On-demand object capture from user-selected (sub)expressions' execution, to which object-centric debugging operations are applied or removed on the fly (e.g., object-centric behavioral instrumentation or breakpoints). | ✓ | ✓ | ✓ | ✓ | ✓ | ✓ | ✓ | ✓ | ✓ | ✓ |
| New gen. debuggers[21] | Instrumentation of methods to record metadata after the execution of an expression and its sub-expressions. Recorded data is used in the debugger to show the last values of executed (sub)expressions. | ✓ | ✓ | ✗ | ✗ | ✓ | ✗ | ✗ | ✓ | ✗ | ✗ |
| Moldable debugger[9, 10] | Evaluation of reflective code conditioning (de)activation of breakpoints before specific message calls/sends within the body of methods. Conditions use metadata reified from the execution context. | ✓ | ✓ | ✗ | ✓ | ✓ | ✗ | ✓ | ✓ | ✓ | ✓ |
| On-demand announcements[65] | On-demand instrumentation of methods to add boilerplate code for an announcement mechanism, without affecting the original base code. | ✓ | ✗ | ✗ | ✗ | ✗ | ✗ | ✗ | ✓ | ✓ | ✗ |
| Post-mortem exec. analysis[64] | Snapshot of objects' internal properties by instrumenting code that changes objects' structure (e.g., all assignments of a variable). | ✓ | ✓ | ✗ | ✗ | ✓ | ✗ | ✗ | ✓ | ✓ | ✗ |
| Interactive feature analysis[19] | Dynamic instrumentation of all methods of the system to produce an execution trace for a dedicated execution. | ✓ | ✗ | ✗ | ✗ | ✓ | ✗ | ✓ | ✓ | ✗ | ✓ |
| Dynamic Aspects[56] | AOP implementation using metalink's meta-objects as advices, and annotating specific AST nodes in methods to model pointcuts. Aspects' execution is controlled using reifications from the execution context. | ✓ | ✗ | ✗ | ✗ | ✓ | ✗ | ✓ | ✓ | ✓ | ✓ |
| Unified reflective API[39] | Alternative API for the reflective features provided by Reflectivity. | ✓ | ✓ | ✓ | ✓ | ✓ | ✗ | ✓ | ✓ | ✓ | ✓ |

*Or *unanticipated*: when instrumentation has not been anticipated in the code before run time.

conditions which result in the (de)activation of breakpoints (i). Conditions use reifications of the execution context (r) (e.g., the receiver of a message).

- *On-the-fly* **object-centric debuggers [11].** These debuggers aim at debugging specific objects of interest without restarting the program. Developers select where to insert the debugging behavior in the source code, and the debugger automatically maps that selection to its corresponding AST node (s). Objects of interest are dynamically captured by dedicated metalinks installed before or after specific operations (cont), such as variable read or write. Object-centric metalinks, associated with meta-objects implementing debugging behavior (i), are installed *on-the-fly* (du) on those captured objects (o). The debugging behavior is dynamically (de)activated through conditions (cond) configured in object-centric metalinks. An object acquires or looses debugging behavior through these metalink conditions, which check local state of the program at run time using user-selected reifications (r).

### 5.3 Limitations

From the study of projects presented in this paper, we relate problems and limitations encountered by users of Reflectivity, how they coped with it and the impact it had on the evolution of Reflectivity. In addition, we describe limitations known by the authors and their knowledge of the library.

**Modifying code disrupts metalinks.** Modifying the source code of methods triggers their recompilation, and results in the loss of installed metalinks in those methods. This happens because metalinks annotate the AST of methods, which changes with the source code. If the AST changes, it is not possible to automatically find on which nodes the lost metalinks should be reinstalled. It has been proposed to warn developers before each recompilation of a method with metalinks [56] and let them decide what to do. Another proposition is to control the presence of metalinks on specific AST nodes like variable accesses [64]. That solution provides means to install metalinks on all accesses to specified named variable. After each recompilation, the AST is checked and for each variable access the metalink is reinstalled. It inspired an API, which has been integrated in Reflectivity. This API is already in use in some projects [12, 14].

**Performance overhead.** The impact of Reflectivity on performance is two-fold: there is an overhead due to the metalink mechanism, and an overhead due to the metalink installation. Some benchmarking were reported [9, 10, 11, 39, 50], but the measured overhead also involved performance loss due to the project using Reflectivity as a backend. Some experiments also reported that the overhead is significantly higher when data is reified from the execution context, and that the more reifications are requested, the higher is the overhead.

Some developers consider that the possible cost in terms of performance overhead can be acceptable compared to the benefit obtained by the solution [11, 50]. It becomes the responsibility of the user to ponder whether that slowdown is acceptable or not, although it does not solve the problem which should be tackled.

To better characterize the impact of using the current release of Reflectivity, we summarize in the following a basic evaluation of the performance overhead due to the use of Reflectivity, and the execution cost of installing metalinks on all methods in a set of classes. Detailed results are available in appendix C.

*Instrumentation Cost.* We measured the instrumentation cost with metalinks of message sends and instance variable read/write. We compared these measurements with their counterpart instrumentation performed from the base level with code insertion. Worst case scenario is for message sends, when calling a meta-object and reifying all data needed to execute the send, which led to a $\approx 75.1\,\%$ slowdown. With less reifications, this overhead diminished to $\approx 16.3\,\%$, which confirms that requesting data from the execution context increases the overhead.

*Installation Cost.* We compared the cost of recompiling an entire set of more than 12000 methods with the installation of a metalink on the same set of methods. Even the worst case scenario here is around 26 % faster than recompilation.

It should be noted, though, that a goal for Reflectivity is to work on an unmodified Pharo virtual machine. We discuss future work about the possibility of changing the virtual machine in Section 6.

**Object-centric reflection.**  In the original Reflectivity implementation, object-centric reflection was limited to filter out objects with conditions before executing a metalink. To cope with this limitation, users extended Reflectivity with object-centric metalinks [11]. Metalinks can now be explicitly installed on specific objects, which in that case are the only objects affected by these object-specific metalinks. Other instances of the same class do not execute metalink related code anymore, nor conditions to control metalink scoping. On basic experiments, it showed a speed-up of 20 % compared to the condition filtering on large sets of objects [11]. This work has been integrated into the official distribution of Reflectivity (section 3.4).

**Limitations inherited from Pharo**   We inherit two limitations from Pharo: thread safety of reflective change and hot code update of methods on the stack.

*Limited thread safety.* Reflective operations that modify the structure, like adding or removing methods or classes, are not thread safe in Pharo (or Smalltalk). For Reflectivity, we inherit this property: installation of metalinks is not thread safe, but once they are installed, the resulting code can be executed safely. The client has to make sure that metalink installation does not interfere, by e.g. using a monitor to coordinate changes from multiple threads.

*Non-applicability to methods on the stack.* Methods already on the execution stack cannot be modified. So if metalinks are installed on such methods, they will only take effect for next calls to those methods. This makes it impossible to add metalinks on methods with an infinite loop that are running.

Both problems need to be solved not just for Reflectivity, but for Pharo in general. We discuss more about this problem in Section 6.

## 6 Future Work

The introduction of Reflectivity in Pharo and its adoption by many developers and researchers showed that it gave an answer to the need for unanticipated, sub-method and partial behavioral reflection. In this section, we propose future directions that will lead to new experimentation and an improved version of Reflectivity.

**Integration with the VM.** We want to explore how Reflectivity can be improved if we give up on the principle that we do not need any modification at the level of the virtual machine. We plan to explore how the Pharo VM can be improved to make Reflectivity more efficient. We think that the JIT could inline metalink calls and thus speed up execution performance. In addition, reifications are dynamically recovered from the execution stack, which is manipulated at run time and thus induce an overhead. We think that some of this information could be provided by the VM instead, reducing the overhead.

**Better sub-method model.** The current implementation of Reflectivity prioritizes practicality in the sense that we wanted it to run it on an unmodified virtual machine. In addition, it was important to keep the implementation a modular addition to Pharo. We therefore do not replace the current reflective model of code (`CompiledMethod`), but we add the AST in addition.

Now that Reflectivity is shipped with Pharo by default, it is interesting to explore what can be done to have one shared model of methods that Reflectivity then could use directly. This would allow us to remove the AST cache and thus speed up instrumentation performance. In addition, we want to explore how to use transformations on this new model to speed up reification. In combination with getting reifications from the VM, an improved AST compilation model with transformations would provide reifications in a static way instead of dynamically manipulating the stack to retrieve them.

Finally, we plan to experiment with non-destructive AST transformations for text to AST compilation: we do not need to re-create the whole AST of a method (and loose all metalinks). Instead the editor could just update the existing AST for every keystroke.

**Thread safety and hot code update** We currently inherit two problems from the implementation platform: thread safety and hot code update of methods that are on the stack. We decided to not solve these problems just for Reflectivity, but instead solve them for Pharo in general. All reflective operations (e.g. changing classes) need to be thread safe. All kinds of code update should provide a way to update methods on the stack. Some research has been done in these directions for Pharo [42, 63], which will be used as the starting point for future research.

**Metalinks on all structural elements.** The Reflectivity metalinks are annotations on the structure of methods in the form of the AST. But there is nothing inherent in the metalink idea that binds it to the AST. In the future we want to extend the concept of metalink to any structural reflective entity: classes, packages. In first experiments, we have added support to annotate first class variable definitions, which is now available in Pharo 8.

**MetaLinks on Data.**  *Metalinks* are only concerned with structural entities of the language: a MetaLink is put, e.g., on the AST node for an instance variable access. But one could imagine the same idea to be applied, in general, to any node in an object graph. For example, if a metalink is set on such a node, it might execute when this object's graph is touched by execution.

**More tools that use metalinks: dynamic information everywhere.**  With Reflectivity we can easily reflect on run-time events in a very fine-grained way. We want to explore if we can enhance the existing development tools to leverage run-time data much more. In addition, we need to make the IDE metalink aware: we need to be able to visualize where metalinks are installed and provide navigation and tool support everywhere. An example of this is a debugger that can optionally step into the meta execution.

## 7 Conclusion

We presented Reflectivity, our reflective library for sub-method, partial behavioral reflection. We introduced the concepts of Reflectivity, its API and the abstractions it provides to the developer. We illustrated how these abstractions are used in Pharo to implement breakpoints through examples.

Reflectivity has been integrated in Pharo since 2008 and provides a playground and a basis for many works integrated inside or outside the Pharo platform. Our investigation of Reflectivity's usage from the literature shows that it attracted many developers to experiment and build new projects on top of it. Through the study of 21 research projects, we analyzed how developers used Reflectivity, how its different features were combined to provide customized reflection capabilities, and its limitations.

Our analysis shows that Reflectivity is used for projects of very different kind, such as building new reflective libraries, code instrumentation, execution analysis, and debugging at run time. Most of these works make a definite use of the properties of Reflectivity, namely behavioral reflection at the sub-method level, in an unanticipated and non-intrusive way. It has proven to be a practical API for expressing and dynamically installing fine-grained reflective behavior (at the level of the AST).

Reflectivity is currently a stable and mature API which lives and updates with the upgrades of Pharo, its current home system. The genericity of its concepts opens new opportunities as future and experimental work. Reflectivity could affect structural elements and data instead of just annotating the AST. This, with a better integration to specialized virtual machines, could open a great field for enhancing new development and debugging tools with efficient, fine-grained run-time information and meta-behavior.

**Acknowledgements** We would like to thanks Stéphane Ducasse for his precious help and feedback during the whole writing process. The following works is supported by I-Site ERC-Generator Multi project 2018-2022. We gratefully acknowledge the financial support of the Métropole Européenne de Lille.

## A   Lists of reifications available in Reflectivity

■ **Table 3**  Lists of reifications depending on AST node types.

| Reification | Target AST Nodes | Description |
| --- | --- | --- |
| #arguments | {*message, method, block*} | Arguments from the node on which the link is installed |
| #class | {*any*} | Class of the object which is executing the code |
| #receiver | {*message, method*} | The receiver of a message |
| #entity | {*any*} | The structural entity on which the link is installed on |
| #link | {*any*} | The metalink itself |
| #method | {*any*} | The method in which the metalink is installed |
| #originalMethod | {*any*} | The method that contains the AST node just before the metalink is installed. |
| #name | {*variable, assignment*} | The name of the variable being read or written |
| #index | {*variable, assignment*} | The index of the indexed instance variable being read or written |
| #newValue | {*variable, assignment*} | The new value written in a variable, when there is a value change |
| #node | {*any*} | The AST node on which the link is installed on |
| #object | {*any*} | The object executing code in which the link is installed |
| #operation | {*any*} | An executable wrapper around the node on which the metalink is installed |
| #selector | {*message, method*} | The selector of a message send or of a method |
| #sender | {*message, method*} | The sender of the method currently being executed |
| #context | {*any*} | The current execution context |
| #value | {*variable, assignment, message, return*} | The current value of a variable read or write, or resulting from an expression evaluation |
| #variable | {*any*} | A variable (global or slot) |

## B  A Look to the Past

Reflectivity is actively used in research projects. From the literature, we collected 21 works using Reflectivity to build novel approaches and/or solutions for answering research or technical problems. We summarize each of these projects — *PhD* thesis, journal, conference and workshop papers — and how they use Reflectivity from a general perspective.

### B.1  Building reflective libraries

The API and abstractions offered by Reflectivity have been used to build reflective libraries or language extensions.

**Unified API to control reflective operations.**  Controlling reflection is a challenging task [39]. The paper presents five dimensions of meta-level control from related literature and proposes a unified API.

**Contribution**  Papoulias, Denker, Ducasse, and Fabresse proposed a meta-object protocol to control reflective operations of Pharo, through a generalized and unified API [39] using a new entity, the reflectogram. When a reflective operation is triggered, a jump to the meta-level is performed and reified. This reification holds contextual information about the meta-level jump and provides fine control over the execution and the result of the reflective operation.

**Reflectivity usage**  All the facilities proposed by this meta-object protocol are implemented on top of Reflectivity. Metalinks are used as means to provide reifications of the execution context, to delegate control and behavior, and dynamically control the (de)activation of the meta-behavior.

**Implementing Aspect-Oriented Programming.**  AOP [28] is a programming paradigm, with which developers can inject cross-cutting concerns in the program. Cross-cutting behavior is encapsulated in language constructs named *advices*, which are injected at key points in the base code named *pointcuts*. The code injection is called *weaving*, and produces programs with an instrumented control flow.

**Contribution**  Strauss introduced AOP for *Squeak* on top of partial behavioral reflection [56]. Reflectivity has been extended to build an implementation of AOP named *Dynamic Aspects*, integrated with existing concepts of Smalltalk.

**Reflectivity usage**  Based on prior work in *Reflex* [58], cross-cutting metalinks are used to implement *Aspects*. The *pointcut* is defined by a special method, that makes use of Smalltalk capabilities to compute the AST nodes where the metalink should be *weaved*. The cross-cutting behavior is directly written in Smalltalk, while the arguments that are passed to the behavior are represented using Reflectivity reifications.

### B.2 Code instrumentation

Reflectivity has been used as a light and flexible code adaptation mechanism in various projects over the years. These projects are presented below.

**Unified structural and behavioral reflection.** In all systems, software needs to evolve. In the case of systems that cannot be stopped, a mechanism is required to alter the software behavior at run time in a very specific and fine-grained level.

**Contribution** Ressia, Renggli, Gîrba, and Nierstrasz used a meta-object protocol to model objects' behavior. They used Reflectivity as a convenient way of refining the behavior of some meta-objects for run-time evolution [44]. Their approach, *Albedo*, is a unified approach to structural and behavioral reflection providing a range of reflective features that can be used to evolve applications at run time.

**Reflectivity usage** *Albedo* relies on Reflectivity to add unanticipated behavior to the running application. In this context, updating an existing behavior is just a matter of installing metalinks on existing methods. The authors illustrated their run-time evolution approach by adding a new feature to a running application, and by dynamically modifying the method lookup of their meta-object protocol.

**Observing and profiling IDE interactions.** Integrated Development Environments (IDE) are heavily used by developers for their daily tasks. Analyzing how developers use their IDE can help them to improve their practices and ease their development process.

**Contribution** *DF2low* is a tool which automatically observes and records user interactions in the Pharo development environment [35]. Recorded interaction events are available to external tools for analysis, for example to provide developers with visualizations of their usage of the development environment's user interface.

**Reflectivity usage** Reflectivity is used for profiling meta-level events issued from user interaction in the Pharo development environment. Metalinks are installed on methods implementing environment related to reflective operations (such as inspect). Those metalinks trigger DF2low profiling behavior with contextual information related to the executed reflective operation.

**Dynamic synchronization.** Synchronization code is required for multi-threaded software. Such code is often mixed with functional code, and it is defined in a static way. Therefore, it cannot be changed at run time to adapt to different situations (e.g., a data race).

**Contribution** Ressia and Nierstrasz introduced a Dynamic Synchronisation System (DSS) which adapts object to different run-time situations [43].

**Reflectivity usage** The authors created an high level API for expressing synchronization definitions based on Reflectivity. The API uses the AST rewriting capabilities of Reflectivity, and reflects at a more conceptual level when dealing with synchronization. The unanticipated code instrumentation proposed by Reflectivity introduces a flexible way of adapting the code at run time, as the synchronization code can be easily modified or removed regarding the situation.

**Augmenting IDEs with Run-time Information for Software Maintenance.** Integrated Development Environments (IDE) are nowadays a very common tool for software development. They are used by almost all developers and, in Smalltalk, they are at the heart of the system. One of the biggest challenges when an IDE is conceived or improved is to adapt it to the user.

**Contribution** Röthlisberger improved IDE user experience by first gathering information about how the users interact with the IDE, then by enhancing it with run-time information and adapting its behavior/features [47, 49, 51].

**Reflectivity usage** The information gathering is performed using Reflectivity. Non-intrusive and temporary code instrumentation allows the authors' tool to dynamically alter the base behavior without modifying its source code. In this scenario, Reflectivity is the main foundation on which the IDE enhancer is built.

### B.3 Debugging

Unanticipated, flexible and non-intrusive code injection is appealing when it comes to debugging. The following papers present work exploiting these capabilities enabled by Reflectivity through unanticipated sub-method and partial behavioral reflection.

**Breakpoint support for Pharo.** The Smalltalk standard way of adding breakpoints in a program is by manually inserting self halt statements in the source code. This presents several downsides [1]: it pollutes the source code, it interferes with the versioning of the code and it can trigger infinite recursions when the insertion site in the source code is poorly chosen.

**Contribution** A Breakpoint model have been implemented and integrated in Pharo [1]. This is today the current breakpoint implementation in latest Pharo distributions.

**Reflectivity usage** A Breakpoint class is used to manage metalinks and their configuration. Instead of manual insertion in the code, a contextual menu has been added to the class browser, from which instances of Breakpoint can be put on the selected source code. The self halt instruction is called from the metalink, and is not inserted in the source code anymore. Infinite recursions are solved by configuring the metalink not to trigger itself recursively.

**Support for out-of-place debugging.** Traditional debugging implies executing debugging instructions in the same execution environment as the debugged program. This can cause *debugging interference*, which might slow down or interrupt the program, or cause unwanted or hidden side-effects (e.g., when evaluating debugging code).

**Contribution** *IDRA* is a remote debugger for distributed Pharo applications [33]. It consists of a master debugger connected to workers, i.e., machines or devices, on which execute instances of the program. When one of these instances encounters an exception, this exception and its execution stack are serialized and transmitted to the master debugger which enqueues them. Developers can debug each exception locally, fix the problem, and deploy the fixed code on all workers.

Sub-method, partial behavioral reflection with Reflectivity

**Reflectivity usage** Reflectivity is used to locally reproduce remote file access in a debugging session. Instead of executing the original file access code, a metalink calls a proxy which performs a remote access to the real file. The debugger can locally re-execute methods with file accesses, by continuing reading the original remote file.

**Unanticipated debugging of non-stoppable applications.** For some applications that cannot be stopped, such as cyber-physical systems, web services or long-running simulations, debugging has to happen *on-the-fly* without restarting the program.

**Contribution** The *Debug-Scopes* debugger [11] is an object-centric debugger. Developers can dynamically express and apply debugging behavior to specific objects without restarting the program. The debugger provides facilities to find objects of interest [12, 14], to express behavioral variations for these specific objects and to express conditions for the (de)activation of the debugging behavior [13].

**Reflectivity usage** The debugger uses Reflectivity to dynamically instrument the control flow of programs with metalinks, and takes advantage of their non-intrusive nature. Metalinks are used to monitor strategic parts of the control flow of the running program, to dynamically gather objects and (de)activate debugging behavior. Object-centric metalinks are installed *on-the-fly* to apply the debugging behavior to the gathered objects.

**New generation debuggers.** Debugging is a challenging activity, which requires adequate tools and techniques to tackle the most difficult problems.

**Contribution** Dupriez, Polito, and Ducasse [21] explore the challenges of today's debugging, and try to foresee new kind of tools to ease the investigation of difficult bugs. They analyze existing debugging tools, and highlight their limitations in regards to the debugging challenges they identified.

**Reflectivity usage** Reflectivity is used for feature design and experiments around debugging. The authors use metalinks as a base instrumentation tool to implement prototypes of what they call new generation debuggers. For example, metalinks are used to enhance conditional breakpoints with additional reified information, or to instrument all AST nodes of a given method to record execution results.

**The Moldable Debugger.** Mainstream debuggers provide low-level, generic operations (e.g., single stepping, logging, breakpoints, etc.), while developers often need domain specific debugging operations and views. Moreover, extending or building *ad-hoc* debugging tools is a complex and costly task.

**Contribution** The Moldable Debugger [9, 10] is a framework to build domain-specific debuggers. It allows developers to adapt the debugging tools to the specificity of their application domain, by providing facilities for building dedicated debugging operations and views.

**Reflectivity usage** Metalinks are used as non-intrusive code-centric instrumentation to control breakpoints activation before methods calls and message sends. Instrumentation ensures that specific conditions are met before halting the execution. For instance, when receiving a message, the instrumentation triggered by the metalink can check if the receiver of the message is a specific object, and halt the execution accordingly.

### B.4 Visualization and execution analysis

On top of debugging features, software analysis plays an important role in software development. Reflectivity has been used to develop instrumentation tools for non-intrusive software analysis and visualizations.

**On Demand Annoucement Mechanism.** The OpenPonk modeling platform is built for developing tools for conceptual modeling like model transformation or DSLs [65]. Moreover, OpenPonk allows the users to create and add plugin to the platform. This platform heavily relies on meta-modeling and model manipulation, which implies a lot of instrumentation of the model code to react on model changes. The plugin system should not directly modify the OpenPonk meta-models, and should gain access to very fine grained information at run time, which is not possible without a very deep modification of the meta-models' base code in order to introduce this mechanism and could introduce overhead.

**Contribution** The authors added an API in OpenPonk which adds a simple way of dealing with change listening.

**Reflectivity usage** OpenPonk uses Reflectivity to define code extension for announcing changes to the model. These extensions can be installed to the desired method on demand without affecting the original code base. The authors implemented an abstraction based on Reflectivity, providing users with a powerful mechanism for on-demand model and meta-model behavior adaptation.

**Post-mortem structural analysis of programs.** Understanding how objects' state and inter-connection evolve during run time is of critical importance when debugging, maintaining and evolving programs. Yet, much information can be implicit due to the dynamic nature of a program execution, and easily missed by developers.

**Contribution** Uhnák and Pergl proposed an approach and a tool to visualize and ease the post-mortem analysis of run-time object structure [64]. The tool selectively captures the execution of a program and exposes its structural properties through live visualizations, in which users can navigate.

**Reflectivity usage** Metalinks are used as the underlying instrumentation mechanism to trigger snapshots of objects and connect them to their domain model. A toolkit[4] based on Reflectivity has been developed to experiment with pervasive instrumentation of the program, for example to target all assignments of a variable and to install persistent metalinks on the corresponding AST nodes.

---

[4] https://github.com/peteruhnak/metalinks-toolkit.

**Incremental and Interactive Analysis of Features.** Many maintenance tasks are usually expressed in terms of user features. Consequently, it is important to establish a correspondence between a feature and the source code it is related to.

**Contribution** Denker, Ressia, Greevy, and Nierstrasz [19] provided a model at run time of features called *live feature analysis* that can establish a map of the different part of the system involved in a user feature.

**Reflectivity usage** The approach uses Reflectivity to automatically annotate whole methods' AST that are used by a dedicated functionality. The system produces traces giving an accurate view of which part of the analyzed system a feature is involved in. It is possible to incrementally enhance feature representation by applying different scenarios to specific features. Reflectivity provides a convenient way for *on-the-fly* annotation of the analyzed software.

## C  An analysis of the performance impact of Reflectivity

In the following, we provide a basic evaluation of the performance overhead due to the use of Reflectivity, and of the execution cost of installing metalinks. Experiments were performed on a MacBook Pro (13-inch, 2018), 2.3 GHz Intel Core i5, memory: 16 GB 2133 MHz LPDDR3. Reproducible experiments for Pharo 7 are available online.[5]

### C.1  Instrumentation cost

We performed a simple benchmark to measure the cost of intrumentation for three entities: message send, instance variable read and write. We compare for these the base case with two instrumentations: calling a method without arguments and calling a method with all the arguments reified that we would need to do the operation on the meta-level. We have thus three cases:

1. No metalink: The reference time.
2. Call a 0-arg empty method on a meta-object (no reifications).
3. Reify all information needed: object, selector, argument for the message send, name and object for the variable access.

We used the benchmark functionality integrated with Pharo, which executes code for five seconds then reports the number of executions of the benchmarked code per second. We ran each benchmark three times. table 4 shows the results. The lower is the number of executions, the slower is the execution.

For variable access, the benchmarked method does both an assignment and a read, this makes the code better comparable, as a method just reading a variable (a getter) would be highly optimized by the virtual machine.

We see that instrumentation results in some overhead. One reason is that the JIT compiler of the current virtual machine cannot inline the message send introduced by

---

[5] https://github.com/MarcusDenker/Reflectivity-Bench.

■ **Table 4** Comparison of the number of executions per second between instrumented and non-instrumented code.

|  | Reference | Call empty method | Overhead |
|---|---|---|---|
| *call meta object, no reification* | | | |
| Message Send | 127 165 905 | 109 269 774 | 16.3 % |
| Var Read | 124 572 050 | 120 963 764 | 2.9 % |
| Var Write | 124 572 050 | 117 053 458 | 6.4 % |
| *call meta object, with reification* | | | |
| Message Send | 127 165 905 | 72 617 729 | 75.1 % |
| Var Read | 124 572 050 | 115 189 016 | 8.1 % |
| Var Write | 124 572 050 | 105 485 492 | 11.8 % |

the metalink. We pay the same price as if we would change the base code and add a call to our meta-object. The reason why the overhead of message sending is quite low is that here multiple bytecodes (for the arguments) are executed in addition to the message send, while for variable read/write we execute a slow message send in addition to the very fast variable access bytecodes. Even the order of bytecode has some impact, as can be seen by the difference of the *var read* vs *var write* overhead.

Additionally, needed data from the base code, for example the creation of the arguments array for a message send, will lead to an additional slowdown. As will, of course, anything that the meta object might execute. This is consistent with results from experiments in works using Reflectivity, which reported that the overhead is significantly higher when data is reified from the execution context, and that the more reifications are requested, the higher is the overhead.

### C.2  Installation cost

We performed a simple benchmark to evaluate the performance overhead induced by metalink installation. We recompiled a set of methods, which would be necessary to instrument them without metalinks, and compared the execution time of that recompilation with the execution time of a trivial metalink installed on all those methods. We chose the Morph class hierarchy, the graphical backend of the Pharo IDE, which represents a total of 12085 methods. Results are reported in table 5. Measurements are performed with and without a hot AST cache. The optionCompileOnLinkInstallation option forces the system to recompile a method immediately after a link is installed to make it comparable to a full recompile.

Installing metalinks is faster than full recompilation, even when recompiling at install time and with an empty AST cache. The reason is that the full compiler provides support for interactive use, while we use a simpler setup for Reflectivity as we can guarantee that no compilation errors will happen (as the code has been compiled

**Sub-method, partial behavioral reflection with Reflectivity**

■ **Table 5** MetaLink installation execution time on a set of 12085 methods, compared to the recompilation (using the default compiler) of the same set of methods without metalinks (in seconds).

|                         | Reference | With metalinks | Speedup |
|-------------------------|-----------|----------------|---------|
| *Compile at install time* |           |                |         |
| Without hot AST cache   | 3.58      | 2.84           | 26 %    |
| With hot AST cache      | 3.58      | 0.92           | 289 %   |

before). With a hot AST cache (for example, installing a second link on the same method), we are more than two times faster.

### C.3 Threats to validity

Benchmarking is not easy as it depend heavily on what is done on the meta-level and what kind of information is reified and passed to the meta-level. It is clear that real applications have to execute *something* on the meta-level to be of practical value. What this is is heavily dependent on the use case.

What we can asses is the *minimal* overhead, *i.e.* the slowdown introduced by Reflectivity itself. We decided therefore to do our benchmarks with an empty method on the meta object. These methods are specially optimized by the VM (no activation record is created). The benchmark thus allows us to assess the minimal slowdown introduced by Reflectivity, allowing us to judge the overhead of Reflectivity itself which is independent of the code that is executed on the meta-level.

**About the authors**

**Steven Costiou** is a permanent researcher in the RMoD team at Inria Lille - Nord Europe. Before that, he worked six years in the industry as a software developer in various areas (defense, aerospace, point-of-sale software). He then did research on unanticipated software adaptation during his PhD at Université de Bretagne Occidentale (France), before becoming a permanent Inria researcher. 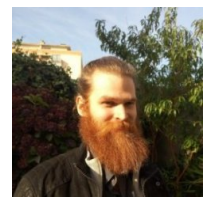

Today, he works on finding new ways of debugging. He is interested in the identification and the study of the properties that programming languages and their infrastructure (i.e., virtual machines) must exhibit to support new debugging features that effectively help debugging. This research spans different topics: reflection and meta-programming, object-centric instrumentation, dynamic software adaptation, dynamic languages and virtual machines. steven.costiou@inria.fr.

**Vincent Aranega** is associate professor at the University of Lille Nord Europe. He received a PhD from University of Lille, France, in nov. 2011, working on model driven development. After the PhD, he worked in the industry for six years on real-time modeling in the cloud, before obtaining a position at the University of Lille. His research interests currently focus on object oriented languages, reflective systems and virtual machines. vincent.aranega@inria.fr. 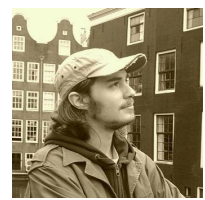

**Marcus Denker** is a permanent researcher at INRIA Lille - Nord Europe. Before, he was a postdoc at the PLEIAD lab/DCC University of Chile and the Software Composition Group, University of Bern. His research focuses on reflection and meta-programming for dynamic languages. He is an active participant in the Squeak and Pharo open source communities for many years. Marcus Denker received a PhD in Computer Science from the University of Bern/Switzerland in 2008 and a Dipl.-Inform. (MSc) from the University of Karlsruhe/Germany in 2004. He co-founded ZWEIDENKER in 2009. marcus.denker@inria.fr. 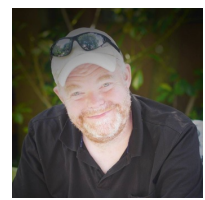